# The Engineering of Optical Conservative Force

Junjie Du,[1,2]* Chi-Hong Yuen,[1,]* Kun Ding,[3] Guiqiang Du,[1]

Zhifang Lin,[4] C. T. Chan,[3] & Jack Ng[1,5]#

[1]Department of Physics, Hong Kong Baptist University

[2]Quantum Institute for Light and Atoms, Department of Physics,

East China Normal University, Shanghai 200062, China

[3]Department of Physics and Institute for Advanced Studies,

The Hong Kong University of Science and Technology

[4]State Key Laboratory of Surface Physics, Key Laboratory of Micro and Nano

Photonic Structures (MOE), and Department of Physics, Fudan University

[5]Institute of Computational and Theoretical Studies, Hong Kong Baptist University

*These authors contributed equally.

#Corresponding E-mail: jacktfng@hkbu.edu.hk

**Abstract**

**Optical forces have been fruitfully applied in a broad variety of areas that not only span the traditional scientific fields such as physics, chemistry, and biology, but also in more applied fields. It is customary and useful to split the optical force into the (conservative) gradient force and the (non-conservative) scattering and**

**absorption force. These forces are different in attributes. The ability to tailor them will open great potential in fundamental optics and practical applications. Here, we present an analytical and a numerical approach to calculate these forces, and, with these tools, we create a fairly general class of 2D conservative optical force field. In general, particles immersed in an optical force do not obey equilibrium statistical mechanics, making the analysis complicated. With conservative forces, these issues are resolved.**

Small particles immersed in a fluid (such as colloids in water) will exhibit Brownian motion due to the random bombardment by the fluid molecules. If the particles are simultaneously illuminated by an intense laser, their motions can be strongly modified and controlled, opening up new applications.[1,2,3,4,5,6,7,8,9,10] Tools like optical tweezers, periodic arrays of optical traps,[11,12,13] or optical lattices, have been used extensively to produce "potential energy landscapes."[14,15,16,17] Nonetheless, recent advancement clearly proves the nonconservative nature of optical forces[18,19,20,21,22,23,24] and for this reason, optical forces do not produce true "potential energy landscapes". Incident light flows in one direction but not the other, this induces nonconservative scattering force that invalidates a potential energy approach. The subtle distinction between conservative and nonconservative forces is that conservative/nonconservative forces can/cannot be derived from a potential energy and the work done by the field is path independent/dependent. Consequently, they are of different attribute and have different applications. For a conservative force, the textbook conservative classical mechanics and equilibrium statistical mechanics may be applied, which is in general significantly simpler than its nonconservative analog.[25,26] Furthermore, a colloid in a conservative field can also be used to simulate other systems where experiments are more difficult[27,28].

It is highly desirable to generate a conservative force field. Nevertheless, in the past, while the total optical force can be calculated[29], people have not yet succeeded in separately calculating the conservative force, called the gradient force $\mathbf{F}_g$ (where $\nabla \times \mathbf{F}_g = \mathbf{0}$) and the non-conservative force, called the scattering and absorption force $\mathbf{F}_k$ (where $\nabla \cdot \mathbf{F}_k = 0$) independently, except for the limiting cases of particle being small[30] or large[32] compare to the wavelength, but not the most experimentally accessible micro-particles.**Error! Bookmark not defined.**,[30,31,32] We note that some attempts to calculate these forces approximately have been made.[33,34] In principles, this could be achieved by using the Helmholtz theorem[35], but this is numerically challenging and physically non-transparent. Probably due to these reasons, Helmholtz theorem has never been implemented in optical micromanipulation.

**Results**

This letter presents a recipe to calculate and engineer the gradient force $\mathbf{F}_g$ and the scattering and absorption force $\mathbf{F}_k$ acting on a spherical object (lossy or transparent). The concept of $\mathbf{F}_g$ and $\mathbf{F}_k$ are of crucial importance in optical manipulation and has been applied in numerous literatures for interpretation and understanding,[36] but their true nature are not known before. Helmholtz theorem states that the total optical force (or any vector field) can be written as $\mathbf{F} = \mathbf{F}_g + \mathbf{F}_k$, where $\mathbf{F}_g = -\nabla U$, $\mathbf{F}_k = \nabla \times \mathbf{g}$.[35] Clearly, $\nabla \times \mathbf{F}_g = \mathbf{0}$ and $\nabla \cdot \mathbf{F}_k = 0$. By Stokes' Theorem, $\nabla \times \mathbf{F}_g = \mathbf{0}$ indicates that the work done on the particle by the field is path independent and therefore a scalar potential can be defined. By optical Earnshaw theorem[30], $\nabla \cdot \mathbf{F}_k = 0$ indicates that $\mathbf{F}_k$ alone cannot confine or trap a particle. To illustrate the strength of our approach, we shall generate optical force field with either $\mathbf{F}_g = \mathbf{0}$ or $\mathbf{F}_k = \mathbf{0}$, as many applications in optical micromanipulation rely on either one. Generally speaking, $\mathbf{F}_g$ and $\mathbf{F}_k$ are, respectively, responsible for optical trapping[31] and transportation[37]. These forces have different attributes and the ability to tailor them may result in qualitative improvements in optical trapping.

Based on a previously derived multipolar expression for optical forces acting on a spherical particle[18], we derived in the supporting information the analytical expression of $\mathbf{F}_g$ and $\mathbf{F}_k$ for the first few multipoles:

$$\begin{aligned}
\mathbf{F}_g &= \frac{\alpha'}{4}\nabla|\mathbf{E}_{in}|^2 + \frac{\beta'}{4}\nabla|\mathbf{B}_{in}|^2 + \frac{\gamma'}{4}\mathrm{Re}\left\{\nabla\nabla\mathbf{E}_{in}^{*}:(\nabla\mathbf{E}_{in} + \nabla\mathbf{E}_{in}^{T})\right\} \\
&+ \frac{\gamma'_m}{4}\mathrm{Re}\left\{\nabla\nabla\mathbf{B}_{in}^{*}:(\nabla\mathbf{B}_{in} + \nabla\mathbf{B}_{in}^{T})\right\} + \frac{\Omega'}{12}\mathrm{Re}\left\{\nabla\nabla\nabla\mathbf{E}_{in}^{*} \vdots sym(\nabla\nabla\mathbf{E}_{in})\right\} + O(k^9 a^9), \\
\mathbf{F}_k &= -\frac{1}{2}\alpha''\mathrm{Im}\left\{\nabla\mathbf{E}_{in}^{*}\cdot\mathbf{E}_{in}\right\} - \frac{k^4}{12\pi\varepsilon_0 c}\alpha'\beta'\mathrm{Re}\left\{\mathbf{E}_{in}\times\mathbf{B}_{in}^{*}\right\} \\
&- \frac{k^5}{40\pi\varepsilon_0}\alpha'\gamma'\mathrm{Im}\left\{(\nabla\mathbf{E}_{in} + \nabla\mathbf{E}_{in}^{T})\cdot\mathbf{E}_{in}^{*}\right\} + O(k^9 a^9).
\end{aligned} \quad (1)$$

where $\alpha'$, $\beta'$, $\gamma'$, $\gamma'_m$, $\Omega'$ and $\alpha''$ are the multipole moments obtainable from Mie theory and are tabulated in the supplementary information, $\mathbf{E}_{in}$ and $\mathbf{B}_{in}$ are the arbitrary incident electromagnetic field, and $sym(\nabla\nabla\mathbf{E})$ denotes all permutation of $(\nabla\nabla\mathbf{E})_{ijk}$. Eq. (1) goes beyond the previous dipolar theory and reaches into the Mie regime. Nevertheless, the particles involved in experiments often have sizes beyond the validity of Eq. (1), due to the truncation in the multipole expansion. The real advantage of Eq. (1) lies in its transparent physics and insight, rather than its ability to compute the numerical values of $\mathbf{F}_g$ and $\mathbf{F}_k$. An independent efficient numerical method based on fast Fourier transform is developed to treat particles with arbitrary sizes:

$$\mathbf{F}_g(\mathbf{x}) = \int \frac{\mathbf{q} \cdot [\mathbf{q} \cdot F(\mathbf{q})]/q^2}{(2\pi)^{3/2}} e^{i\mathbf{q} \cdot \mathbf{x}} d^3\mathbf{q}, \quad \mathbf{F}_k(\mathbf{x}) = \int \frac{[\mathbf{q} \times F(\mathbf{q})] \times \mathbf{q}/q^2}{(2\pi)^{3/2}} e^{i\mathbf{q} \cdot \mathbf{x}} d^3\mathbf{q}, \quad (2)$$

where **x** is the coordinate of the sphere center, $F(\mathbf{q}) = (2\pi)^{-3/2} \int \mathbf{F}(\mathbf{x}) e^{-i\mathbf{q} \cdot \mathbf{x}} d^3x$ is the Fourier transform of the total optical force and $\mathbf{F}(\mathbf{x})$ is the total optical force, which can be calculated by using the Maxwell stress tensor and the generalized Mie theory. It can be readily verified that in both Eq. (1) and(2), $\mathbf{F} = \mathbf{F}_g + \mathbf{F}_k$ while $\nabla \times \mathbf{F}_g = \mathbf{0}$ and $\nabla \cdot \mathbf{F}_k = 0$. Eq. (2) is a practical numerical approach to calculate $\mathbf{F}_g$ and $\mathbf{F}_k$ for particle of arbitrary size. Eq. (2) also works for non-spherical particles, but one then has to take care of the particle orientation.

Figure 1 plotted $\mathbf{F}_g$ and $\mathbf{F}_k$ for a 300 nm diameter dielectric spherical particle illuminated by an *x*-polarized fundamental Gaussian beam. The total optical force needed in Eq. (2) is computed by the highly accurate Mie theory. The strongly focused incident field is computed by the vector generalization of the Debye's integral (see supporting information)[38,39], which is known to generate results that can be directly compare with experiment. The forces are calculated analytically using Eq. (1) and numerically using Eq. (2). As far as for the 300nm-diameter dielectric particle is concerned, excellent agreement is achieved, which validates both analytic and numerical approaches. In Fig. 2, we also plotted $\mathbf{F}_g$ and $\mathbf{F}_k$ using Eq. (2), for the widely employed linearly polarized fundamental Gaussian beam, which is also known as the standard optical tweezers. To obtain converged calculation for the strongly focused Gaussian beam, the unit cell for Fourier transformation is chosen such that the forces near the edges are at least two to three orders of magnitude smaller than that of the center. We remark that our approach can also be used in other nonconservative force field, such as the acoustic force.

We now return to the generation of a force field characterized by $\mathbf{F}_k = \mathbf{0}$. A careful inspection of Eq. (1) surprisingly reveals that $\mathbf{F}_k = \mathbf{0} + O(ka)^9$ if $\text{Im}\{\mathbf{E}_{in}\} = \mathbf{0}$, which can be fulfilled by an incident standing wave. In other words, a standing wave can generate a conservative force field for particles with diameter less than roughly half a wavelength. Fig. 3(a), (b), and (c) plotted, respectively, the potential energy $U$ (where $\mathbf{F}_g = -\nabla U$), $|\mathbf{F}_g|$, and $|\mathbf{F}_k|$ for a 1 micron diameter particle illuminated by a standing wave generated by interfering four plane waves (each with an intensity of $10^4$ W/cm$^2$). Clearly, a conservative periodic potential is generated, since $|\mathbf{F}_g| \gg |\mathbf{F}_k|$. To our surprise, $|\mathbf{F}_k|$ in Fig. 3(c) is smaller than the numerical noise. Also, a one micron diameter particle is beyond the validity of Eq. (1), and therefore the observed conservative force is unexpected. We attempted to calculate $\mathbf{F}_k$ and $\mathbf{F}_g$ for particles as large as 5 microns in diameter (data not shown), $\mathbf{F}_k$ is still below the numerical errors. Apparently, reasoning beyond Eq. (1) is needed to explain the observed phenomenon. We analytically proved in the supporting information that for a spherical particle in a 2D (i.e. all incident wave vectors lie on the $z$=0 plane) TE or TM standing wave, the optical force is conservative, just as shown in Fig. 3. We remark that the vanishing of the scattering and absorption force is a consequence of symmetry. We can still generate a conservative force even for a lossy particle.

For comparison, the forces acting on a one micron diameter particle when illuminated by three plane waves are plotted in Fig. 4. These three plane waves do not form a standing wave, but their incident momentum do cancel each other completely, i.e. they have the same amplitude and $\sum_{i=1}^{3} \mathbf{k}_i = \mathbf{0}$. When the waves are coherent, both $\mathbf{F}_k$ and $\mathbf{F}_g$ are non-zero due to interference. The force field is clearly nonconservative, as the maximum value of $\mathbf{F}_k$ is actually greater than that of $\mathbf{F}_g$. This highlights the importance of having a standing wave. In fact, an odd number of plane waves will not be able to create a purely conservative optical force, as they cannot produce a standing wave.

**Discussion**

It is previously known that under appropriate conditions, a standing wave can exert a conservative force on a dipolar particle. Here, our coherent and systematic approach provides a sufficient condition for achieving a purely conservative force field with periodic or localized field pattern for spherical particles of any size.

We now turn to the generation of pure $\mathbf{F}_k$. The only way to generate a transverse nonconservative force with zero gradient force is to use orthogonal plane waves, because, non-orthogonal plane waves (or Fourier components) will unavoidably interfere to produce gradient forces. Such a setup is too simple to be engineered for applications. It is more effective to combine $\mathbf{F}_k$ and $\mathbf{F}_g$ in transportation application.

In summary, we devised an analytical and a numerical approach to calculate and engineer the gradient force and scattering and absorption force for the experimentally accessible micro-particles. The profile of these forces associated with the optical tweezers is presented. A sufficient condition to induce a conservative force is presented. This will enable more detailed analysis and precise control on optical micromanipulation. As an example, we provided a recipe to generate or tailor a rather general 2D conservative force field using a TE or TM standing wave. This will allow us to create a truly conservative force field in 2D, paving the way to mimic a wide variety of phenomena in equilibrium statistical mechanics using optical micromanipulation system.

**Methods**

*Focused Gaussian beam calculated by generalized vector Debye integral*

The strongly focused Gaussian trapping beam is modeled by using the highly accurate generalized vector Debye integral. In short, an incident unfocused laser beam is illuminated on a high numerical aperture objective lens. Since the lens is macroscopic in size, the focusing problem can be treated using geometrical optics with negligible errors. Then the vector Debye integral is used to connect the geometrical optics solution to the focal region. Such an approach is known to yield results that can be quantitatively compare with experiments[13,38,39].

*Optical force calculation by generalized Mie theory and Maxwell stress tensor*

The total optical force has time independent components and time dependent components. The latter oscillates at a rate of twice the frequency of light. At such a fast frequency, the particle cannot response to it, and therefore it suffices to consider the time-independent component, or the time averaged optical force, which has been referred as the optical force throughout the text.

The (time averaged) optical force is computed by:

$$\mathbf{F}_{total} = \oiint_{\text{particle surface}} \mathbf{T} \cdot d\mathbf{a}, \quad (3)$$

where

$$\mathbf{T} = \tfrac{1}{2}\varepsilon_0 \mathbf{E}\mathbf{E}^* + \tfrac{1}{2}\mu_0 \mathbf{H}\mathbf{H}^* - \tfrac{1}{4}\varepsilon_0 |\mathbf{E}|^2 \ddot{\mathbf{I}} - \tfrac{1}{4}\mu_0 |\mathbf{H}|^2 \ddot{\mathbf{I}} \quad (4)$$

is the time averaged Maxwell stress tensor. The electromagnetic field needed in the Maxwell stress tensor are total field, which is the incident field (strongly focused trapping beam or plane waves) plus the scattered field from the particle. The latter is computed by using the generalized Mie theory.

**Acknowledgements**

The work is supported by Hong Kong RGC through GRF 603312, ECS 209913, and AoE/P-02/12. ZFL was supported by NNSFC through 11174059.

**Author contributions**

J.D. wrote the FFT program and produced the major results. C.H.Y. assisted the FFT calculation and wrote the proof. K.D. assisted the analytical calculation. G.D. contributed in discussion. Z.F.L. provided the calculation direction. C.T.C. guides and oversees the research. J.N. initiated the project, invented the methodologies, and oversaw the research.

**Additional information**

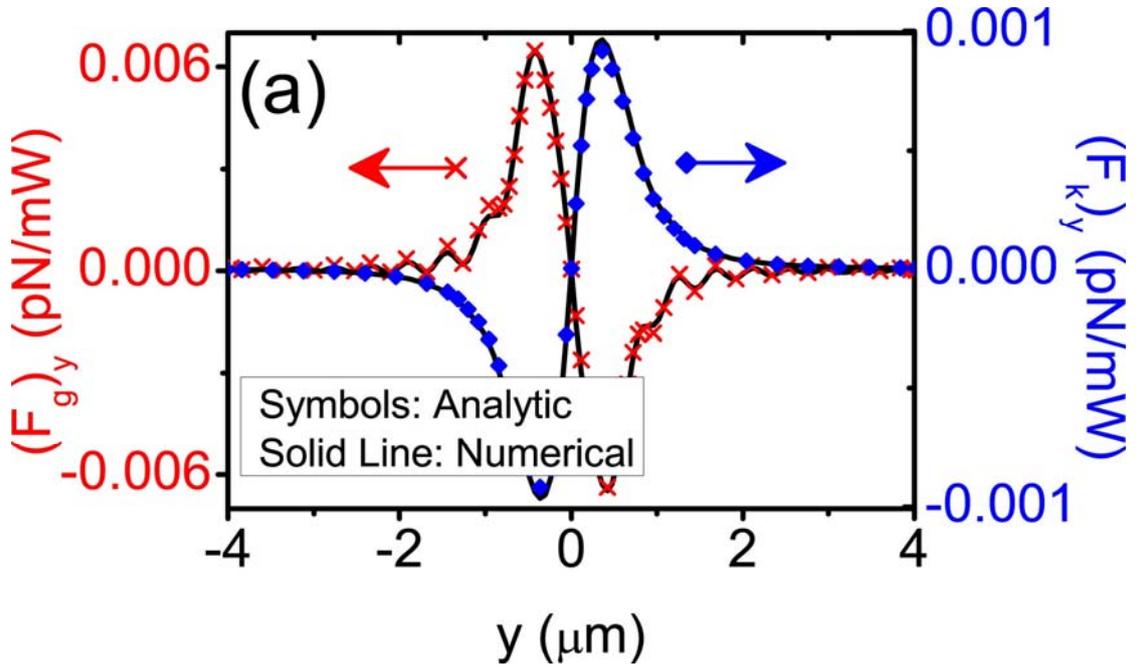

**Fig. 1**: Excellent agreement of analytic and numerical approaches is obtained. Gradient force (red) and scattering force (blue) calculated by the analytical expression Eq. (1.3) (symbols) and numerical approach Eq. (1.4) (solid lines), respectively. The 300 nm diameter particle is immersed in water. The wavelength is 1064 nm. The refractive indices of the particle and water are 1.59 and 1.33, respectively.

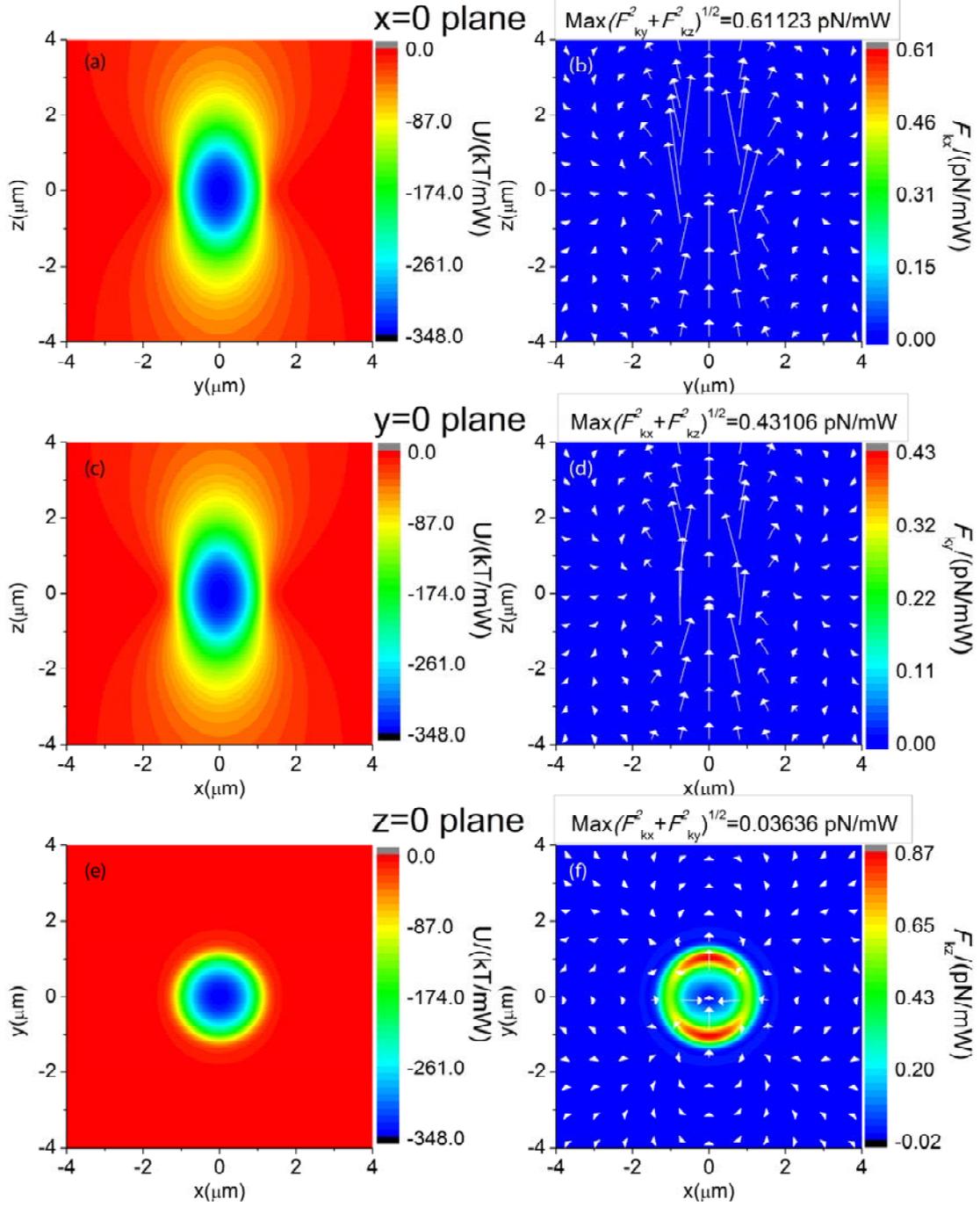

**Fig. 2:** The incident beam is a strongly focused, *z*-propagating, and *x*-polarized fundamental Gaussian beam (standard optical tweezers) in water ($\varepsilon_{water} = 1.33$). The Numerical Aperture N.A. is 1.3 and the filling factor is 1. **Left** Potential energy *U* of the gradient force for a 1 micron diameter polystyrene particle, where $\mathbf{F}_g = -\nabla U$. **Right** Scattering force. Arrows indicate the direction and magnitude of force in logarithmic

scale. Panels (a)-(b), (c)-(d), and (e)-(f) are for the *x*=0, *y*=0, and z=0 planes, respectively.

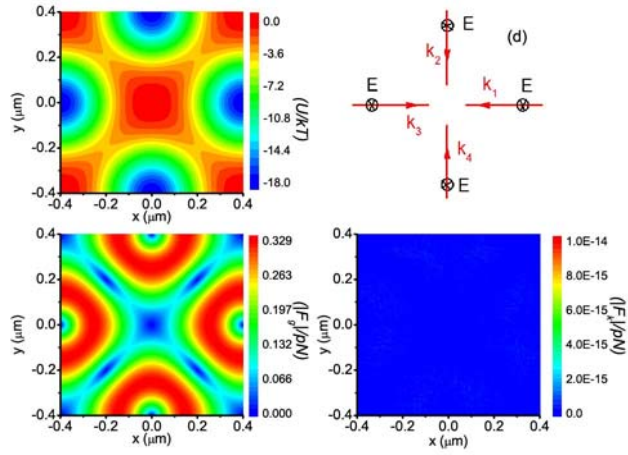

**Fig. 3:** Conservative optical force acting on a 1 micron diameter particle in water illuminated by a standing wave depicted in part **(d)**. **(a)** Potential Energy *U*. **(b)** $|\mathbf{F}_g|$. **(c)** $|\mathbf{F}_k|$. Clearly $|\mathbf{F}_k| \approx 0$, therefore the force is conservative. **(d)** Schematic illustration for the configuration of the incident plane waves.

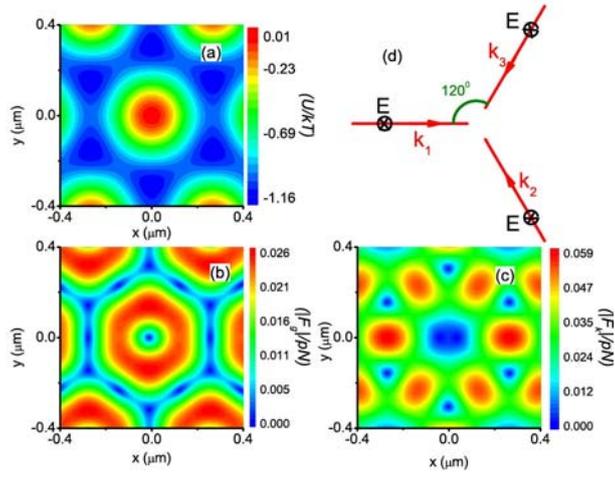

**Fig. 4:** Nonconservative optical force when the incident wave is not a standing wave. **(a)** Potential Energy $U$. **(b)** $|\mathbf{F}_g|$. Clearly $|\mathbf{F}_k| \neq 0$, therefore the force is nonconservative. **(c)** $|\mathbf{F}_k|$. **(d)** Schematic illustration for the configuration of the incident plane waves.